\documentclass[10pt,a4paper]{article}

\usepackage{amssymb,amsmath,amsthm}
\newcommand{\uvec}[1]   {%
  \mathrel{\vbox{\offinterlineskip\ialign{%
    \hfil##\hfil\cr
    $\scriptscriptstyle\hookrightarrow$\cr
    \noalign{\kern .2ex}
    $#1$\cr
}}}}
\usepackage{graphicx}
\usepackage{color}
\usepackage{paralist}
\usepackage{titlesec}
\usepackage{wasysym}
\usepackage[bookmarks=false]{hyperref}
\usepackage{amssymb}
\usepackage{amsmath}
\usepackage{graphicx}
\usepackage{color}
\usepackage{colortbl}
\usepackage{multirow}
\usepackage{multicol}
\usepackage{lipsum}
\usepackage{balance}
\usepackage{algorithmic,algorithm}

\graphicspath{{./}{figures/}} 

\newcommand{\cancel}[1]   {\ignorespaces}

\begin{document}

\newtheorem{definition}{Definition}
\newtheorem{theorem}{Theorem}[section]
\newtheorem{lemma}{Lemma}[section]
\newtheorem{corollary}{Corollary}[section]
\newtheorem{claim}{Claim}[section]
\newtheorem{proposition}{Proposition}[section]
\newtheorem{fact}{Fact}[section]
\newtheorem{observation}{Observation}[section]

\newbox\ProofSym
\setbox\ProofSym=\hbox{%
\unitlength=0.18ex%
\begin{picture}(10,10)
\put(0,0){\framebox(9,9){}}
\put(0,3){\framebox(6,6){}}
\end{picture}}

\def\proof{\par\noindent {\it Proof.}\ }
\def\endproof{\hfill\hbox{$\quad \copy\ProofSym$}\break}
\renewcommand{\qedsymbol}{\copy\ProofSym}

\numberwithin{equation}{section}

%\newenvironment{emromani}
 %{\renewcommand{\theenumi}{\roman{enumi}}
   %\renewcommand{\labelenumi}{{\em (\theenumi)}}
   %\begin{enumerate}}
  %{\end{enumerate}}

%%%%%%%%%%%%%%%%%%%%%%%%%%%%%%%%%%%%%%%%%%%%%%%%%%%%%%%%%%%%%%%%%%%%%%%%%
%                                                                       %
% Preambles. DO NOT ERASE THEM. Change to suite your particular purpose.%
%                                                                       %
%%%%%%%%%%%%%%%%%%%%%%%%%%%%%%%%%%%%%%%%%%%%%%%%%%%%%%%%%%%%%%%%%%%%%%%%%

\title{Denoising a Point Cloud for Surface Reconstruction}
\author{Siu-Wing~Cheng \quad\quad Man-Kit Lau \\
Department of Computer Science and Engineering \\
HKUST, Hong Kong \\
\{scheng,lmkaa\}@ust.hk}

\date{}

\maketitle

\begin{abstract}

Surface reconstruction from an unorganized point cloud is an important problem
due to its widespread applications.  White noise, possibly clustered outliers,
and noisy perturbation may be generated when a point cloud is sampled from a
surface.  Most existing methods handle limited amount of noise.  We develop a
method to denoise a point cloud so that the users can run their surface
reconstruction codes or perform other analyses afterwards.  
%Our method consists of a novel octree decomposition, a statistical analysis
%routine, and a novel variant of meshless Laplacian smoothing.  The input data
%points are first grouped into clusters by using our octree decomposition, and 
%
%
%
% 
%clusters the data points using a novel octree decomposition.  We also develop
%a simple statisitcal analysis and a novel variant of meshless Laplacian
%smoothing to remove outliers and noise from the clustered data points.
%
%Finally, we apply the Robust Cocone algorithm to reconstruct the surface from
%the denoised point cloud followed by some postprocessing on the reconstructed
%surface for constructing the boundaries and further smoothing. Our algorithm
%applies a variant of the standard octree structure to manipulate the points
%and this makes our algorithm very efficient. 
%
%The experimental results show that our method is fast, and it can produce a
%good reconstruction even though the noise level is very high.  The data sets
%that we experimented with include some raw data contaminated by artificial
%noise, white noise and clusters of outliers. 
Our experiments demonstrate that our method is computationally efficient and it
has significantly better noise handling ability than several existing surface
reconstruction codes.
%We compare our method with several surface reconstruction codes.  %with Robust
%Cocone, the Adaptive Moving Least-Squares (AMLS) algorithm, and %the PoissonPU
%algorithm, and Not only is our method computationally efficient, the denoised
%point set also enables reliable reconstruction in the presence of noise and
%outliers.

\end{abstract}

\section{Introduction}

%%%%%%%%%%%%%%%%%%%%%%%%%%%%%%%%%%%%%%%%%%%%%%%%%%%%%%%%%%%%%%%%%%%%%%%%%
%                                                                       %
% The following are a number of example sections/subsections.           %
%                                                                       %
%%%%%%%%%%%%%%%%%%%%%%%%%%%%%%%%%%%%%%%%%%%%%%%%%%%%%%%%%%%%%%%%%%%%%%%%%

Surface reconstruction from a point cloud is an important problem due to its
widespread applications.  Algorithms for reconstructing from noise-free data
have been developed.  Some of them
\cite{paper:Amenta98b,paper:Amenta98,paper:Amenta00,paper:Amenta00PowerCrust}
produce provably good approximations when the data satisfies certain sampling
conditions.
%Amenta and Bern \cite{paper:Amenta98} and Amenta, Choi, Dey, and Leekha
%\cite{paper:Amenta00} developed the first provably correct surface
%reconstruction algorithms. When the sample is dense enough, the output
%approximates the unknown surfaces well in the sense that their normal
%deviation is small, their Hausdorff distance is small, and they are
%homeomorphic. More research works on surface reconstruction can be found in
%[].
However, it is a challenge to handle white noise, outliers, and noisy
perturbation that may be generated when sampling a point cloud from an 
unknown surface.  Outliers may even be structured and clustered (e.g.~sample 
points in the planar background when an object is scanned).

%
%For example, the measurement error made by a 3D laser scanner may produce
%noisy perturbation and outliers (including points scanned from unwanted
%background).  Several surface reconstruction codes~\cite{paper:Bernardini99,
%paper:RobustCocone, paper:Kazhdan06} can handle noise to a limited extent, and
%the reconstruction quality degrades rapidly for large noisy perburbations.
%
%Moreover, they do not cater for white noise and outliers. 
%
%
%However, most of the surface reconstruction algorithms are sensitive to noise
%while the unorganized point cloud data is typically generated by scanning the
%surfaces of real objects via a 3D laser scanner which may contain noise and
%outliers from various sources.  In the presence of noise and outliers, surface
%reconstruction becomes more challenging. 
%

A popular strategy for denoising a point cloud is to define a smooth
surface function using the point cloud (e.g.~\cite{paper:Dey05, paper:Huang09,
paper:Kol05, paper:MLSLevin03, paper:Mederos03, paper:Nagai09, paper:Schall05,
paper:Shen04}).  If a denoised point set is desired, the input data points can
be projected iteratively to the zero set of the surface function, or a meshing
algorithm can be applied.  The efficiency depends heavily on whether 
the surface function can be evaluated very quickly because many evaluations are
needed.  Most of these surface
functions require accurate oriented surface normals at the data points.
%
%, but noise and outlier filtering seem to be necessary in order to obtain
%accurate normals.  
It has been reported that reconstruction algorithms may be brittle if they
depend on accurate surface normal estimation~\cite{paper:Mullen10}.
%%Inaccurate normal estimation may lead to a poor approximation of the unknown
%%surface poorly.  If a surface mesh is desired after constructing the surface
%function, it can be computed by a meshing algorithm which often evaluates the
%surface function at many places.  Evaluating the surface function at a point
%often requries scanning at least the input data points in the local
%neighborhood.  Depending on the meshing algorithm, repeated evaluations may be
%as time-consuming as scanning the input point cloud multiple times.

%
%
%evaluating the the surface function may be expensive if it is not a
%polynomial because the surface function is evaluated for numerous times during
%the running of the algorithm.  

In this paper, we take a different approach to denoise the point cloud,
assuming that the data points are sampled densely from a smooth surface with or
without boundaries.  Our goal is to produce a denoised point set on which the
users can run their surface reconstruction codes or perform other analyses.  In
our experiments, we 
%give an example that different reconstruction codes can be run on the output
%of our denoising algorithm, although we mainly use 
run Robust Cocone~\cite{paper:RobustCocone} on the point sets denoised by our
method.  %to compare our reconstruction quality with other methods.  
Our denoising code can be downloaded from the webpage for this
project~\cite{proj}. 
%{\footnotesize {\tt http://www.cse.ust.hk/faculty/scheng/denoise.html}}.
%{\footnotesize {\tt https://ihome.ust.hk/\~{}lmkaa/cgi-bin/DenoisingSurfRecon/DenoisingSurfRecon\_projectPage.html}}

%Our method is designed to handle white noise, outliers, and large noisy
%perturbations.  
In our experiments, the number of white noise points and possibly clustered
outliers can be more than 100\% of the number of data points, and the noisy
perturbation can be as large as 2\% of the bounding box diameter of the point
cloud.   The amount of outliers and noise pose a serious challenge to several
existing surface reconstruction codes as indicated in our experiments.  
%This paper focuses on denoising the point cloud.

Our denoising method consists of three stages.  First, we use a new octree
decomposition to cluster the data points.  The largest cluster contains the
data points around the unknown surface; therefore, by extracting the largest
cluster, we can remove white noise and outliers.  If there are $k > 1$
surfaces, we work with the $k$ largest clusters instead.  Second, we perform a
simple statistical analysis on the octree boxes to remove data points with
relatively large noisy perturbation.  Third, we use the remaining points to
guide the construction of a sparser point set, and we develop a meshless
Laplace smoothing procedure to denoise this new point set.  Unlike previous
approaches, we do not estimate surface normals, or perform surface fitting, or
determine the inside/outside of the unknown surface, which helps to make our
method simple and fast.

%After denoising, the user can apply existing methods to finish the
%reconstruction or perform other anlayses. 

%
%and it does not fit the data points to any local or global smooth surface
%which can be time-consuming. 

%There are several novel features in our algorithm. 

Our octree decomposition is based on a previous work with a
collaborator~\cite{paper:Cheng12} for surface reconstruction from clean data.
We extend it so that the octree-induced clustering facilitates the removal of
outliers and white noise.  
%The octree also allows fast access from a data point to other data points in
%its proximity.

We find it desirable to divide the removal of noisy perturbation of the data
points into two steps, the first step for removing relatively large perturbation
and the second step for removing any smaller perturbation left.  Otherwise,
if we apply the meshless Laplacian smoothing right after removing the outliers
and white noise, the ``weighted averages'' will drift away from the
unknown surface.  

There is a novel feature in our meshless Laplacian smoothing step.  Instead of
smoothing the input data points, we construct a sparser point set and smooth
these points.  The reason is that it is computationally inefficient to shift an
input data point based on all other input data points in its neighborhood.  We
use the input data points to guide the construction of a sparser point set, but
it is undesirable to apply Laplacian smoothing yet because if a point is
surrounded by other points, the unevenness in the local sampling will cause the
point to drift in the direction of higher local density during smoothing.  Poor
output may be produced.
%away from the unkonown surface.  
Instead, we divide the neighborhood of a point in this sparser point set into
groups, and we only pick one representative point from each group.  This gives
an even sparser point set to which smoothing is applied.

\subsection{Previous Work}

Previous work perform the noise and outlier filtering together with the
surface reconstruction.  Therefore, we briefly survey the related surface
reconstruction work.

\subsubsection{Point Cloud Denoising}

%There are several methods in the literature for denoising a point cloud. A
%popular one is called \emph{moving least-squares} (MLS).  This method is
%designed for interpolating scattered data.  Several variants of MLS have been
%proposed for approximating the unknown surface from an unorganized point
%cloud~\cite{paper:Alexa01, paper:Amenta04, paper:PMLSAnalysis, paper:Dey05,

There are several variants of \emph{moving least squares} (MLS) for
approximating the unknown surface under noise and outliers.
%
%~\cite{paper:Alexa01, paper:Amenta04, paper:PMLSAnalysis, paper:Dey05,
%paper:Kol05, paper:Levin98, paper:MLSLevin03, paper:Mederos03, paper:Shen04}.  
%
Shen et al.~\cite{paper:Shen04} defined the \emph{implicit moving least-squares}
(IMLS) surface for a soup of polygons with their oriented normals as
constraints.  Kolluri~\cite{paper:Kol05} adapted IMLS to a point cloud in which 
every data point is tagged with an oriented surface normal.  He proved that the
approximation is good if a uniform sampling condition is satisfied.  Dey and
Sun \cite{paper:Dey05} proposed the \emph{adaptive moving least-squares} (AMLS)
surface definition which allows the sampling to be non-uniform and sensitive to
the local feature size.  
% 
% 
%\emph{adaptive moving least-squares} (AMLS) surface definition.  the method
%in~\cite{paper:Dey06}.  Dey and Sun prove that AMLS yields a good The
%approximation is provably good  
%
%
%that relaxes the
%uniform sampling requirement to an adaptive one that is sensitive to the local
%feature size.  
%The normals at the data points are obtained by an extension of the method
%in~\cite{paper:Dey06}.  Dey and Sun prove that AMLS yields a good
%approximaiton. 
%
%for denoising point cloud with theoretical guarantees under an adaptive
%sampling condition.  Accurate normal estimation is important for AMLS. 
%
%Dey and Sun extend the method in~\cite{paper:Dey06} to estimate normals for a
%noisy point cloud.  
%
AMLS can handle a small amount of noise in our experiments.
%yield good approximations when the noise is high.  in such cases.
%
%, so the AMLS surface does not approximate the unknown surface well in such
%cases. 

%
%With the
%implicit surface function, their algorithm denoises the points by projecting
%them onto the AMLS surface via the Newton method. From our experiments, the
%normals are not accurate enough for a very noisy point cloud, so the surface
%function does not approximate the real surface well.

%There are some other denoising algorithms. 

Schall, Belyaev and Seidel~\cite{paper:Schall05} defined a likelihood function
for data points associated with oriented surface normals.  Outliers are removed
by thresholding on the function values.  The remaining data points are
projected to local functional maxima to produce a denoised point set.
%Again, estimating normals using noisy points is difficult.  Also, 
%However, structured outliers (e.g.~outliers on the planar background in the raw
%{\sc Dragon} data) may be mistaken for parts of the surface.

Xie, Wang, Hua, Qin and Kaufman~\cite{paper:Xie03} defined a surface function
by blending quadric surfaces defined over different regions.
%.  The quadric surface patch is found from a flatter region first and then
%propagate to its neighbor.  The points are projected onto the corresponding
%quadric surface.  %This approach cannot reconstruct the shape well at narrow
%regions or at %regions where at least two surfaces come close. 
%
%
%
%This approach may not reconstruct the shape well at regions where at least two
%surfaces come close and may give a two-sheets quadric surface at a flat region
%with large noisy points.  the quadric surfaces may be difficult to approximate
%the real surface because 
It seems hard to determine the extent of these regions for a very noisy point
set.  Xie, McDonnell, and Qin~\cite{paper:Xie04} used an octree to classify the
space into regions that are inside/outside the unknown surface, fit a surface
to each octree cell, and blend these surfaces.  Non-uniform sampling and sharp
features are allowed.  Their experiments show that a small amount of noise and
outliers can be handled.
%In contrast, we use a new octree decomposition for clustering the data points
%only instead of classifying every part of the space as inside/outside.

Nagai, Ohtake and Suzuki~\cite{paper:Nagai09} presented the PoissonPU algorithm
for constructing an implicit surface.  
%It is based on partition of unity.  
Surface normals are required at the data points. 
%
%
%
%This implicit surface function requires the normal information of the center
%of each sphere generated from PU. The normal at each center is computed by
%taking the weighted average of the normals at the sample points inside the
%sphere. 
They apply anisotropic Laplacian smoothing to the surface normals to deal with
normal estimation error, while preserving features.  A small amount of noise
and outliers can be handled as shown in their experiments.
%
%The outlier removal works successfully for a small amount of unstructured
%outliers.  In our experiments, structured outliers and large noise
%perturbations pose a problem for PoissonPU.  , and it runs slower on most of
%the clean data than our method.

%
%
%
%
%
%preserving detailed features, 
%
%the
%normal at each center should be blended with the normals at the neighboring
%centers in order to have a better reconstructed surface. Their algorithm
%applies anisotopic Laplacian smoothing to smooth the normals while preserving
%detailed features, where the 1-ring neighbors of each center $c_i$ of the
%sphere $s_i$ used in the Laplacian smoothing is defined to be the center of the
%spheres that intersect with $s_i$. The implicit surface function is updated
%using the smoothed normals, and then the zero-level surface is reconstructed by
%SurfaceNets \cite{paper:Gibson98}. Their experimental results show that this
%algorithm can work with noisy normals. This approach also computes a confidence
%values for each sample point for filtering outliers. 
%
%The outlier removal works
%successfully for a small amount of unstructured outliers but it cannot handle
%structured outliers. 

Mullen, De Goes, Desbrun, Cohen-Steiner and Alliez~\cite{paper:Mullen10}
proposed to impose signs on an unsigned distance function defined for a point
cloud.  The inside/outside test is performed by tracing rays and 
%
%
%a signed distance function to approximate the surface by imposing signs to an
%unsigned distance function in \cite{paper:Chazal10}. 
checking the parity of intersections with an $\epsilon$ band around the zero
set of the distance function.
%the inside/outside sign by counting the parity of intersections with an
%$\epsilon$ band around the zero set of the implicit function.  of times the
%ray intersect with an $\epsilon$ band. For each point, several rays are
%shooted and a confidence value for outlier detection is then defined by the
%number of inside/outside signs obtained from the rays. The surface
%reconstruction from the signed distance function is obtained by Delaunay
%refinement.  A surface mesh is obtained by Delaunay refinement~\cite{book}.
%They can handle noise, boundaries, and unstructured outliers.  The method
%works for noisy point clouds with unstructured outliers and boundaries.  
Structured and clustered outliers may be an issue (e.g. sample points in the
planar background in the Dragon data set) as they may fool the inside/outside
decision. The large amount of outlier clusters in our experiments pose some
difficulties.
%because the number of times that a ray intersect with the $\epsilon$ band may
%be affected by the structured outliers. 

Giraudot, Cohen-Steiner, and Alliez~\cite{paper:Giraudot13} developed a
noise-adaptive distance function that can handle any smooth submanifold of
known dimension, variable noise, and outliers.  They report that some data sets
are challenging for reconstruction algorithms that rely on accurate surface
normals.  The clustered outliers in our experiments seem to pose a challenge to 
their code.
%In their experiments, they can handle an impressive amount of
%variable noise and outliers, but their code has a very high running time.

% can
%be as large as 60\% of the number of true data points. 
%A higher level of outliers can be tolerated by our method. 

Guggeri, Scateni and Pajarola~\cite{paper:Guggeri12} proposed a depth carving
algorithm to reconstruct the unknown surface.
%
%to determine whether a point is inside or outside the unknown surface.  
Their experiments show that some white noise can be handled.  However,
boundaries pose a problem and no result is given on handling noisy
perturbation.

\subsubsection{Triangulation Algorithms}

There are several triangulation algorithms \cite{paper:Amenta00PowerCrust,
paper:Bernardini99, paper:Cocone, paper:Dey01, paper:tightCocone,
paper:RobustCocone, paper:Kazhdan06, paper:Lorensen1987} for surface
reconstruction and some are able to handle a small amount of
noise~\cite{paper:Bernardini99, paper:RobustCocone, paper:Kazhdan06}. 

Lorensen and Cline \cite{paper:Lorensen1987} proposed the seminal marching cube
algorithm which is a fast way to obtain a triangular mesh of the zero set of an
implicit function.  However, it is sensitive to noise and outliers. 
%intersection of the given surface and the cubes.  This approach gives an
%approximation of the surface quickly if the implicit surface approximates the
%unknown surface well. 

Bernardini, Mittleman, Rushmeier, Silva and Taubin \cite{paper:Bernardini99}
proposed a ball-pivoting algorithm for surface reconstruction of a point cloud
by a region growing method.  Oriented surface normals at the data points are required.
%and these normals need to be correctly oriented.  
%The resulting mesh interpolates the input sample points. 

Kazhdan, Bolitho and Hoppe \cite{paper:Kazhdan06} 
%proposed a Poisson surface reconstruction algorithm which 
expressed the surface reconstruction as a Poisson problem.  A 3D indicator
function that best-fits the point cloud is computed.  Then, an appropriate
isosurface is extracted to approximate the unknown surface.  Oriented surface
normals at the data points are required. 
%and the reconstruction quality depends significantly on the accuracy of these
%normals.

%This limits the noise level that this algorithm
%can tolerate. 

Dey and Goswami \cite{paper:RobustCocone} developed the Robust Cocone algorithm 
%which is a Delaunay/Voronoi based surface reconstruction algorithm 
to reconstruct a closed triangular mesh from data points with a small
amount of noise. 
% and is able to handle noisy
%data. 
The mesh consists of an appropriate set of Delaunay triangles induced by the
data points.  It is a provably good approximation if
%has theoretical guarantees if the input point set
the point cloud satisfies certain sampling conditions.
%and the reconstructed surface is shown to be homeomorphic to the real surface. 
%
%In our experiments, the output of Robust Cocone degrades significantly when a
%small amount of noise and outliers are introduced.  

Kolluri, Shewchuk and O'Brien \cite{paper:Kolluri04} proposed the Eigencrust
algorithm.
%that can handle noise and filter outliers.  
After computing the Delaunay tetrahedralization, spectral graph partitioning is
used to determine whether a Delaunay tetrahedron lies inside the
unknown surface.  The reconstruction 
%
%
%
% 
%the Delaunay tetrahedralization and the corresponding Voronoi diagram, and
%then the algorithm uses a variant of spectral graph partitioning to determine
%whether the Voronoi vertices are inside or outside the surface.  The
%reconstructed mesh 
consists of the triangles shared by inside and outside tetrahedra.  Their
experimental results show that the reconstruction can be corrupted when there
are many outliers or the noise level is high.
%high noise level.  a large number of outliers exist or the noise level is
%high.

\label{sec:RobustCocone}

%%%%%%%%%%%%%%%%%%%%%%%%%%%%%%%%%%%%%%%%%%%%%%%%%%%%%%%%%%%%%%%%%%%%%%%%%
%                                                                       %
% An example of a figure. Note how the Figure Number is generated in    %
% the list of figures.                                                  %
%                                                                       %
%%%%%%%%%%%%%%%%%%%%%%%%%%%%%%%%%%%%%%%%%%%%%%%%%%%%%%%%%%%%%%%%%%%%%%%%%

%\begin{figure}[h]
%\caption{Illustration of Church's Thesis}
%\end{figure}

\section{Octree Decomposition}
\label{sec:octree}

%\subsection{Notation and Definitions}
%\label{sec:NotationsAndDefinitions}

Consider an axes-aligned minimum bounding cube $B$ of the input point cloud.
We present a new octree decomposition to
subdivide $B$ into smaller cubes.  

Every cube that we refer to is aligned with the coordinate axes. The
\emph{size} of a cube $x$ is its side length, and we denote it by $\ell_x$.
Every octree node corresponds to a cube, and therefore, we refer to an octree
node as a cube.  %
%For any two octree nodes $x, y$, their boundaries are in contact if $|c_{x_d}
%- c_{y_d}| \leq |\frac{l_x}{2} + \frac{l_y}{2}|$ for all $d$ and $|c_{x_d} -
%c_{y_d}| \geq |\frac{l_x}{2} - \frac{l_y}{2}|$ for at least one $d$, where
%$c_{x_d}$ and $c_{y_d}$ denote the $d$-th coordinate value of $c_x$ and $c_y$.
Two octree nodes are \emph{neighbors} if their interiors are disjoint and their
boundaries are in contact.  Two nodes are \emph{neighbors at the same level} if
they are neighbors and have the same size.  If a cube is split, it is
partitioned by three planes orthogonal to the coordinate axes into eight
smaller cubes of equal size.  

A leaf node is splittable if it satisfies the following
condition, which was proposed by the authors and a
collaborator~\cite{paper:Cheng12} for surface reconstruction from clean data.
%
%Figure \ref{fig:nodeNeighborExample} is an example in 2D case, and let $x$ be
%the node with black boundary. $x$ and the square with gray boundary has no
%boundary in contact and their interiors are overlap, thus they are not
%neighbors. $x$ and the square with red boundary have boundaries in contact but
%their interiors are overlap, so they are not neighbors. $x$ and each of the
%two green squares have boundaries in contact and their interiors are disjoint
%to $x$ but $x$ has larger side length, i.e. $x$ is neighbors with these two
%squares but not neighbors at the same level. $x$ and each of the two blue
%squares have boundaries in contact, their interiors are disjoint and they have
%same side length, i.e. $x$ is same level neighbors with these two squares.
%

\begin{quote}
Let $x$ be a leaf node.  Let $S$ denote the partition of $x$ into $8\times 8 \times 8$ 
disjoint cubes with size $\ell_x/8$.  The node $x$ is  \emph{splittable}
if at least two cubes in $S$ contain some input data points.
\end{quote}

\begin{figure}
\centering
\includegraphics[scale=0.6]{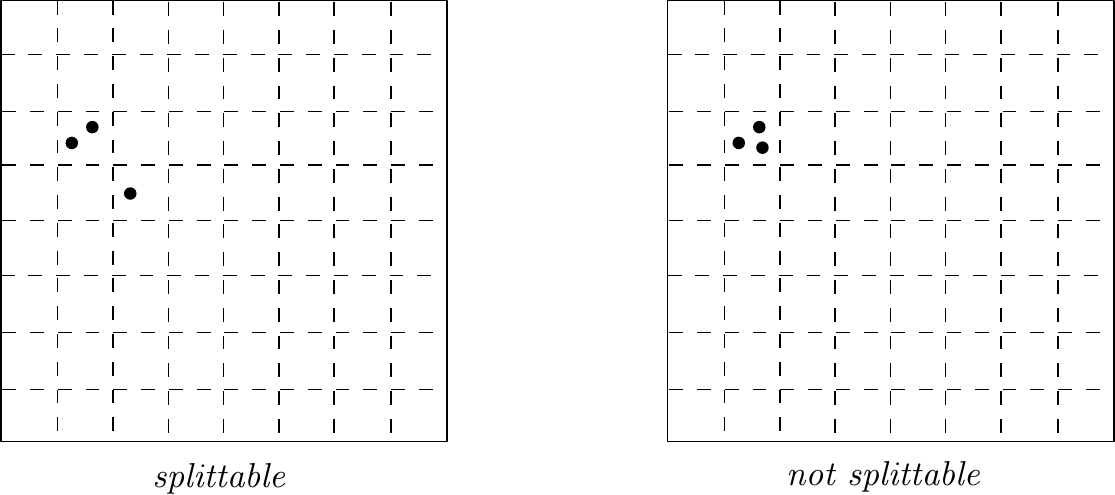}
%\scalebox{0.8}{\includegraphics{figures/algorithm/splittableExample.pdf}}
\caption{2D case example of splittability.}
\label{fig:splittableExample}
\end{figure}

Fig.~\ref{fig:splittableExample} illustrates the idea in 2D. The left
figure shows an octree node $x$ containing $3$ data points.  After dividing
$x$ into $8^2$ square of equal sizes, two of these squares contain some data 
points.  Therefore, $x$ is splittable. The right figure is a non-splittable
node that contains $3$ data points. 
%but only one of the squares formed by the
%dashed lines contains sample points and thus this is not splittable. The
%partition by the dashed lines is only used for determining the whether a node
%is splittable, and the any of the cubes in the partition may not exist in the
%final octree.

For every octree node $x$, we say that $x$ is \emph{empty} if there is no
data point inside $x$.  We call an octree \emph{balanced} if 
neighboring leaf nodes differ in size by at most a factor 2. 

%Given a point set $P$, let $T_P$ denotes the octree constructed from $P$.

%A triangle $t$ is a boundary triangle if an edge of $t$ that is not incident to
%any other triangle. A Delaunay ball of a triangle $t$ is the ball such that the
%three vertices of $t$ lies on the ball.

%\begin{figure}
%\centering
%\includegraphics[scale=0.3]{figures/algorithm/nodeNeighborExample.pdf}
%%\scalebox{0.8}{\includegraphics{figures/algorithm/nodeNeighborExample.pdf}}
%\caption{2D case example of neighbor nodes}
%\label{fig:nodeNeighborExample}
%\end{figure}

The initial octree consists of just the root corresponding to the bounding cube
$B$.  We maintain a queue $Q$ that contains all the current, non-empty leaf
nodes whose splittability have not been checked yet.  Initially, $B$ is the
only item in $Q$.  In a generic step, we dequeue a leaf node $x$ from $Q$ and
calls a procedure {\sc SplitBalance}$(x)$ until $Q$ becomes empty.  We maintain
the invariant that the current octree is balanced after each call of {\sc
SplitBalance}.

In {\sc SplitBalance}$(x)$, we check whether $x$ is splittable, and if so, we
call a procedure {\sc Split}$(x)$.  In {\sc Split}$(x)$, we first partition $x$
into eight equal-sized cubes $z_i$, $i \in [1,8]$, which become the children of
$x$ in the octree, and then we call a procedure {\sc Balance}$(z_i)$ for all $i
\in [1,8]$, and finally we enqueue the leaf nodes $z_i$, $i \in [1,8]$, into
$Q$.  In {\sc Balance}$(z_i)$, we check whether the octree is unbalanced, which
can happen if and only if for some neighboring leaf node $y$ of $z_i$, $\ell_y
= 2\ell_x$ and $\ell_y = 4\ell_{z_i}$.  That is, $x$ and $y$ were just ``in
balance'' and $z_i$ and $y$ are not ``in balance'' after the splitting of $x$.
For every such leaf node $y$, we call {\sc Split}$(y)$ recursively to restore
the balance.  

\vspace{8pt}

\noindent {\bf Remark~1:} The non-empty leaf nodes in the final octree have the
appropriate sizes in the sense that if we select one data point from each
non-empty leaf node, then the selected data points form a locally uniform
sample of the unknown surface,\footnote{This fact is proved
in~\cite{paper:Cheng12} for a slightly different octree decomposition in which
the node splitting and node balancing do not interleave.  That is, all
splittable nodes are split first before the restoration of balance.  
Interleaving node splitting and balancing gives a better practical
performance.} i.e., the local density of the selected
sample points varies smoothly over the unknown surface.

\vspace{8pt}

Let $T$ denote the resulting octree.  Although this octree definition cannot
handle outliers and noise, it serves as a good starting point.  The key
observation is that if the point cloud is clean and it is a fairly uniform
sample of the unknown surface, then most non-empty leaf nodes should have the
same size.  Even in the presence of outliers and noise, the mean
octree cell size $\ell_\mathrm{avg}$ is informative.  

\begin{figure*}
\centering
\includegraphics[scale=0.8]{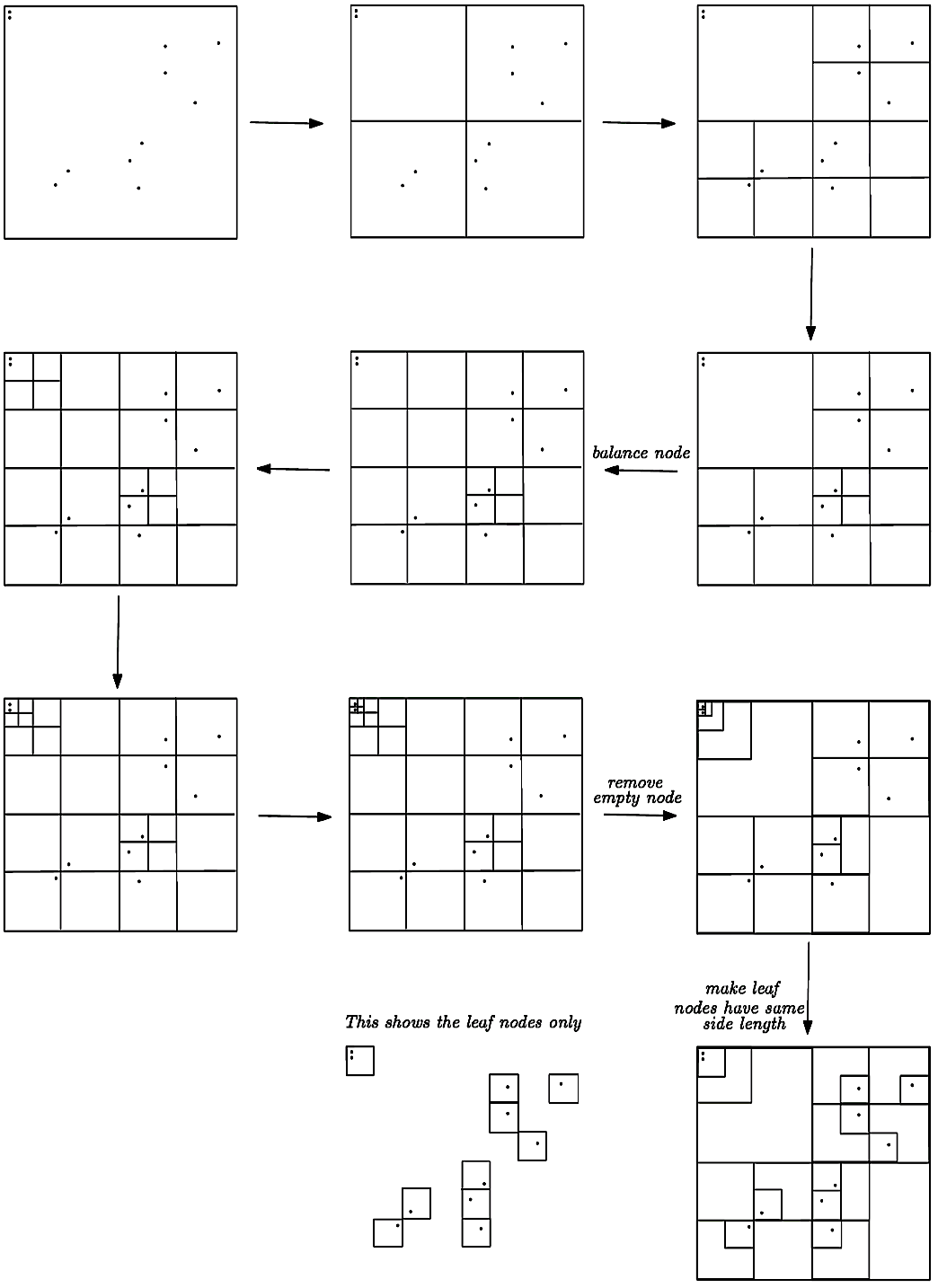}
\caption{2D case example of octree construction.}
\label{fig:octreeConstructionExample}
\end{figure*}

As a result, we use a real-valued parameter $\alpha \geq 1$ and convert the
octree $T$ into another octree such that every leaf node has the same size
$\ell_P$, where $\ell_P \in (\frac{1}{2}\alpha
\ell_{\mathrm{avg}},\alpha\ell_{\mathrm{avg}}]$.  (We set $\alpha = 2$ in
almost all of our experiments.)  This is done as follows.  Take a leaf node $x$
in the current octree.  If $\ell_x > \alpha \ell_{\mathrm{avg}}$, split $x$;
otherwise, if $\ell_x \leq \frac{1}{2}\alpha\ell_{\mathrm{avg}}$, delete $x$.
We repeat the above until $\frac{1}{2}\alpha\ell_{\mathrm{avg}} < \ell_x \leq
\alpha \ell_{\mathrm{avg}}$ for every leaf node.  We use $T_P$ to denote the
final octree obtained.
%
%After building the octree $T_P$ by the above algorithm, each leaf node has the
%same side length $l_P$. The major difference between the octree structure in
%\cite{paper:Cheng12} and our octree structure is we have step $4$ and $5$. We
%refine the octree to make every leaf node has the same side length $l_P$. 
Fig.~\ref{fig:octreeConstructionExample} illustrates the octree construction in
2D.  The fact that every leaf node has the same size $\ell_P$ is useful for our
outlier detection to be discussed in Section~\ref{sec:RemoveOutliers}.

%Our denoising algorithm will need to access from an octree node to its
%neighboring octree nodes of the same size, i.e., these nodes are at the same
%level of the octree.  We faciliate such fast access by storing cross-reference
%pointers among neighboring octree nodes at the same level.

%We need this because our outliers detection algorithm works when the side length difference between any two leaf nodes are small even though the two leaf nodes are not near to each other and thus we simply set all leaf nodes have the same side length. This will be discussed in section \ref{sec:RemoveOutliers}. %Without step $6$, we cannot avoid this situation even though we have the balancing step. For example, consider a sample point which is close to a corner of the root node and all other points are around the opposite corner, then the resulting octree contains some leaf nodes with a large side length difference. However, the balancing step is necessary, since this step can improve the value $l_{avg}$ to represent how dense of the points near the real surface.

\section{Denoising}
\label{sec:denoise}

\subsection{White Noise and Outlier Removal}
\label{sec:RemoveOutliers}

%In practice, our outlier removal approach can sustain certain non-uniformity
%such as some small regions are heavily or not so heavily sampled. For clear
%explanation of this approach, we assume the points close to the real surface
%are uniformly distributed.

\cancel{
We construct an octree $T_P$ for the input point set $P$ by the procedure
mentioned in section \ref{sec:OctreeConstruction} and assume the leaf nodes in
$T_P$ that contain the sample points close to the real surface form a cover to
the real surface and there exists sample points surrounding these leaf nodes.
%\par
}

\begin{figure}
\centering
\includegraphics[scale=0.6]{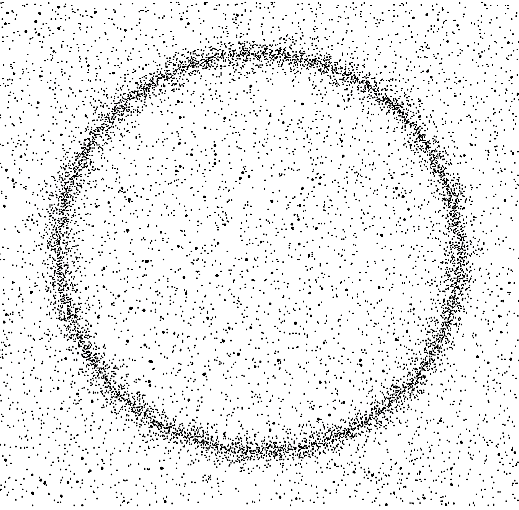}
%\includegraphics[scale=0.45]{figures/algorithm/denseNoiseCircle6287combine.pdf}
%\scalebox{0.8}{\includegraphics{figures/algorithm/denseNoiseCircle6287combine.pdf}}
\caption{Noisy sample of a circle.}
\label{fig:denseSparseCircle}
\end{figure}

Fig.~\ref{fig:denseSparseCircle} shows some noisy points sampled from a
circle surrounded by white noise and outliers.  Our outlier removal procedure
is based on two ideas.  First, the data points near the true surface are denser
than the white noise.  However, outliers may also form dense clusters which
appear as dark, thick dots in Fig.~\ref{fig:denseSparseCircle}.  Our second
idea is that the data points near the true surface occupy a bigger space
than a coincidental cluster of outliers.  We explain below how to turn these
two ideas into efficient procedures.

\cancel{

The data points near the true surface should be denser than the white noise and
occupy larger space than the outliers
%in the sense of human perspective.
as illustrated the 2D case in figure \ref{fig:denseSparseCircle}.
%the points shown in the left is generated by $67832$ points where $62782$
%points around the circle with noise and $5000$ points are added as white noise
%and clusters of outliers. The points shown in the right is generated by $5484$
%points where $484$ points around the circle with noise and $5000$ points are
%added as white noise and clusters of outliers.
In the left figure it is easy to observe that the points are representing a
circle since the points around the circle is much denser than the white noise.
Some cluster of outliers may be denser than the points around the circle, but
the points around the circle forms a cluster which occupy a larger space
compare to any cluster of outliers. In the right figure, it is difficult to see
a circle, since the points around the circle is not much denser than the white
noise. We have designed an approach to remove outliers based on this
observation and our assumption.
}

We construct a graph $G$ in which the vertices are the leaf nodes in $T_P$, and
two vertices in $G$ are connected by an edge if the corresponding leaf nodes
are neighbors at the same level of $T_P$.  We run a breadth-first search on $G$
to obtain its connected components.  Each connected component is a cluster of
data points in $T_P$, and we expect the data points near the true surface to
induce a large connected component.  The white noise points are sparser, and
therefore, a leaf node of $T_P$ that contains white noise points may well be at
distance greater than $\alpha \ell_{\mathrm{avg}} \geq \ell_P$ from other leaf
nodes containing white noise points.  A cluster of outliers may induce a
connected component in $G$.  However, such a connected component has very few
vertices when compared with the connected component induced by the data points
near the true surface.  The same can be said for connected components induced
by white noise points that happen to near each other.  Consequently, 
if $P$ is sampled from a single surface, we can simply extract the data
points in the largest connected component in $G$.  If $P$ is sampled from $k
\geq 1$ surfaces, then we extract the $k$ largest connected components in $G$.
The knowledge of $k$ is a rather weak requirement because the user often knows
how many surfaces are to be reconstructed.  One can also try $k = 1$, check the
output, and then increase $k$ if necessary.

\cancel{

With the above observation and our
assumption, we conclude that the leaf nodes of $T_P$ contain sample points that
are close to the real surface, they should form a connected component in $G$ if
$\alpha$ is sufficiently large where $\alpha$ controls the side length of the
leaf nodes. Therefore, we run a breath-first search on $G$ to obtain the
connected components.

\par
Each leaf nodes have the same side length $l_P$. Thus, the sample points inside a connected component with the largest number of nodes, they form a cluster of points that occupies the largest space compare with the other connected components. If $P$ is representing a single surface, we extract the sample points in the largest connected component be the points that are close to the real surface while the points in other connected components are treated as outliers. If the user would like to reconstruct $k > 1$ surfaces from $P$, one can extract the sample points in the largest $k$ connected components and apply the steps in section \ref{sec:RemoveHighNoise} and \ref{sec:MeshlessLaplacianSmoothing} for each connected component separately.
}

A larger $\alpha$ tolerates a higher non-uniformity in the density of the data
points near the true surface, but fewer outliers and white noise points are
removed.  If $\alpha$ is smaller, more outliers and white noise points are
removed, but the point density near the true surface should be more
uniform. 

\subsection{Filtering Very Noisy Points}
\label{sec:RemoveHighNoise}

\cancel{
After removing the outliers, we obtain a subset $P' \subseteq P$ of sample
points. However, $P'$ may still contain some very noisy points, which makes
denoising difficult. Therefore, we remove the very noisy points in $P'$ based
on a simple statistical analysis.  

\par 
}

Let $P'$ denote the subset of $P$ that we extracted from the largest $k$
connected components of $G$.  When the noisy perturbation is large, some points
in $P'$ may still be quite noisy and relatively far from the true surface.  We
eventually use a variant of meshless Laplacian smoothing to handle the noisy
perturbation; however, according to our experience, the smoothing works less
satisfactorily when some points in $P'$ suffer from large noisy perturbation.
We discuss in this subsection how to remove very noisy points
from $P'$.

\begin{figure}
\centering
{\includegraphics[scale=0.45]{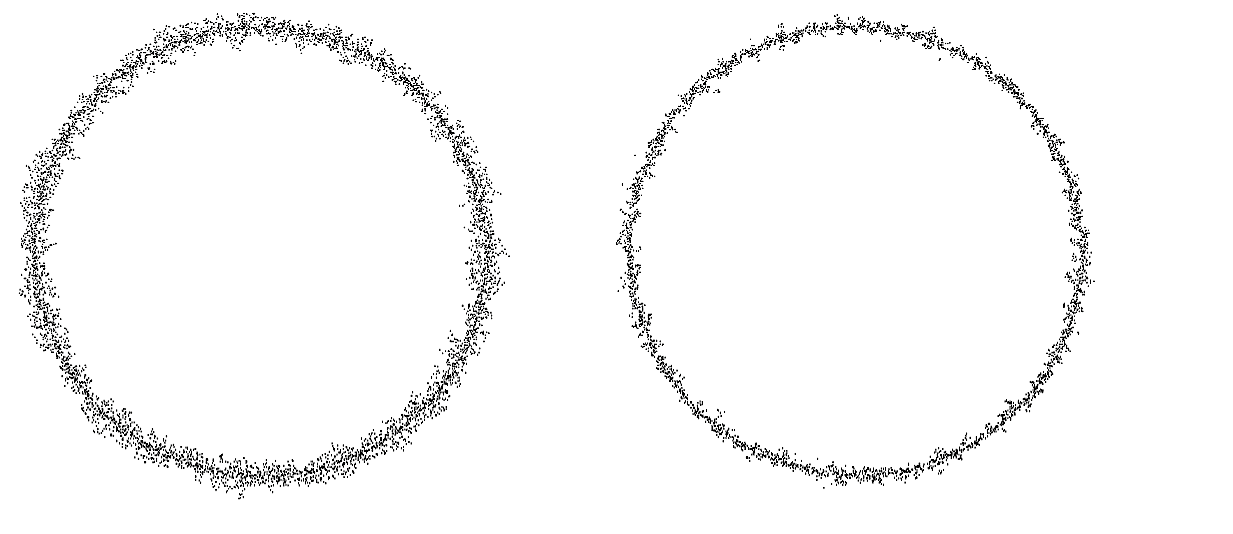}}
%\scalebox{0.8}{\includegraphics{figures/algorithm/removeHighNoise.pdf}}
%\caption{2D case example of outlier removal (left) and filtering very noisy points (right) effect.}
\caption{Left: after removing white noise and outliers.  Right: after removing
the very noisy points.}
\label{fig:removeHighNoise}
\end{figure}

The left image in Fig.~\ref{fig:removeHighNoise} shows the resulting point set
after removing the white noise and outliers in
Fig.~\ref{fig:denseSparseCircle}.  We want to remove the peripheral points
(very noisy points).  These peripheral points have a relatively sparser
neighborhood (i.e., fewer points in the neighborhood).  We introduce an
efficient pruning method based on this observation.

We prune $T_P$ to $T_{P'}$ by deleting the points not in $P'$ and removing the
nodes that become empty afterwards.  
\cancel{ For the points in $P'$, points of low noise should be denser and
closer to the real surface while points of high noise should be sparser and
farther away from the real surface. We make use of the neighborhood size of the
leaf nodes to determine which nodes are closer to the real surface. A classic
way to define the neighborhood size of a point $p$ is to compute the number of
points within a ball centered at $p$ with radius $r$. In this step, we use
another way to define the neighborhood size. For any leaf node $x$ in $T_{P'}$,
we define the neighborhood size of $x$ be the number of sample points inside
the cube centered at the center of $x$ with side length $5l_{P'}$.
%(See figure \ref{fig:neighborhoodSizeForRemoveHighNoise}).
We adopt this way to define the neighborhood size instead of the classic way
because this is more efficient. If each node $x$ keeps the number of points
inside it during the octree construction, we only need to look at a constant
number of nodes to obtain the neighborhood size of $x$.} 
Define the \emph{neighborhood size} of a leaf node $x$ in $T_{P'}$ be the
number of data points in the cube that is centered at the center of $x$ with
side length $5\ell_P$.  Let $n_{\mathrm{avg}}$ denote the average and
$n_{\mathrm{sd}}$ denote the standard deviation of the neighborhood sizes among
all the leaf nodes in $T_{P'}$.  Let $\beta \geq 1$ be a parameter.  (We set
$\beta = 2$ in our experiments.) If $\beta n_{\mathrm{sd}} > n_{\mathrm{avg}}$,
there is still a large variation in neighborhood sizes due to very noisy
points, and therefore, we remove the points in leaf nodes that have
neighborhood sizes at or below the $1$ percentile.  Afterwards, we update
$n_{\mathrm{avg}}$, $n_{\mathrm{sd}}$, and the neighborhood sizes of the nodes
affected.  We repeat the above removal of points until $\beta n_{\mathrm{sd}} \leq
n_{\mathrm{avg}}$.  
\cancel{ Since $n_{\mathrm{sd}}$ decreases and $n_{\mathrm{avg}}$ increases
after each iteration of removal, this procedure will stop before removing the
points of low noise and close to the real surface if $\beta$ is sufficiently
small.
%$n_{sd}$ decreases faster than $n_{avg}$ and thus it will not remove the
points close to the real surface when $\beta$ is sufficiently large.  In
practice, this step only has effect when some points in $P'$ are extremely
noisy. We keep this step so that we can produce a surface when all points in
$P'$ are very noisy.
%\par
Fig.\ref{fig:removeHighNoise} shows the effect of this step on a 2D circle.
The left figure is the result after the outlier removal procedure described in
section \ref{sec:RemoveOutliers} on the point set shown in the left of figure
\ref{fig:denseSparseCircle}. The right figure is the result after this
filtering procedure.  }
In Fig.~\ref{fig:removeHighNoise}, the right image shows the result of
pruning the left image.
%after applying the above filtering to the left image.

\cancel{
After removing the very noisy points, we obtain a
subset $P'' \subseteq P'$
%. We scale the points in $P''$ such that the minimum bounding cube of $P''$
%has side length $2
and construct a new octree $T_{P''}$ for $P''$ by the procedure mentioned in
section \ref{sec:OctreeConstruction}. 
}

\subsection{Meshless Laplacian Smoothing}
\label{sec:MeshlessLaplacianSmoothing}

%In section \ref{sec:LaplacianSmoothing}, we have a brief review to Laplacian
%Smoothing, which is an approach to smooth a given mesh but not for an
%unorganized point cloud. In this section, we present an approach that applies
%the Laplacian Smoothing idea to smooth an unorganized point cloud.

Let $P''$ be the remaining point set after removing from $P$ the white noise,
outliers, and the very noisy points as described previously.  Laplacian
smoothing is a classic algorithm for smoothing a polygonal mesh. It
%developed from the heat diffusion equation and 
works by moving a point $p_i$ iteratively along the direction which is a
weighted average of the vectors from $p_i$ to its 1-ring neighbors.  We use a
variant of this scheme.  In the following, we first introduce our scheme by
showing how it works on $P''$.  Eventually, we construct another point set to
replace $P''$ and run our scheme on this point set instead.

For every point $p_i \in P''$, let $N(p_i)$ denote a set of neighbors of $p_i$
that will be defined later.  For each $p_i \in P''$ and
for each $p_j \in N(p_i)$, define 
\begin{eqnarray*}
d(p_i) & = & \max_{p_j \in N(p_i)} \| p_i-p_j \|,\\ 
w(p_i,p_j) & = & \mathrm{exp}\left(-\| p_i - p_j\|^2/d(p_i)^2\right).
\end{eqnarray*}
Let $\lambda$ be a parameter that controls the step size of moving $p_i$ in
each iteration.  Let $\gamma$ be a parameter so that a point $p_i$ is moved
only if $p_i$ will move over a distance greater than the average distance
between $p_i$ and $N(p_i)$ divided by $\gamma$.  (We set 
$\lambda = 0.25$ and $\gamma = 40$ in our experiments.)
The pseudocode of the 
smoothing process is given below.

\cancel{
$\frac{m_i}{\gamma}$. Repeat
the LaplacianSmoothing$(P'', \lambda, \gamma)$ procedure iteratively until no
point moves or the number of iteration exceeds a predefined threshold.

The major problem to apply Laplacian smoothing on unorganized point cloud
is the absence of the 1-ring neighbors information. Our meshless Laplacian
smoothing approach identifies the 1-ring neighbors from the unorganized point
cloud and then apply the Laplacian smoothing algorithm mentioned above.

In our
algorithm, we apply a modified version of the classic Laplacian Smoothing
algorithm, which is described as follows,
}

\begin{quote}
\noindent {\sc Smooth}$(P'', \lambda, \gamma)$
%\vspace{-.075in}
\begin{enumerate}

\item[1.] For each point $p_i \in P''$, compute 
\[
m(p_i) = \frac{1}{|N(p_i)|}\sum_{p_j \in N(p_i)} \Vert p_i - p_j\Vert.
\]

\item[2.] For each point $p_i \in P''$, compute 
\[
p_i' = p_i + \lambda \cdot \frac{\sum_{j \in N(p_i)} w(p_i,p_j)(p_j - p_i)}{\sum_{j\in N(p_i)} 
w(p_i,p_j)}.
\]

\item[3.] For each point $p_i \in P''$, if $\Vert p_i - p_i'\Vert > m(p_i)/\gamma$, 
then set $p_i = p_i'$.
%Compute $m_{avg} = \frac{1}{|P''|}\sum_{p_i \in P''}\sum_{j \in N_i} \frac{\Vert p_i - p_j\Vert}{|N_i|}$.
%\item[3.]
\end{enumerate}
\end{quote}
\cancel{
where $N_i$ is the set of 1-ring neighbors of $p_i$,
%$w(p_i,p_j)$ is a weight function,
the weight function $w(p_i,p_j) = e^{\frac{-\Vert p_i - p_j\Vert^2}{d^2}}$
where $d$ is the largest distance between $p_i$ and its 1-ring neighbors,
$\lambda$ is a parameter which controls the step size of moving $p_i$ in each
iteration and $\gamma$ is a parameter which allows a point $p_i$ to be moved
only if $p_i$ will move over a distance of $\frac{m_i}{\gamma}$. 
}
{\sc Smooth}$(P'', \lambda, \gamma)$ is called repeatedly until no point in
$P''$ moves anymore or until the number of calls reaches a predefined
threshold.  
%We set the threshold to be ... in our experiments.

\cancel{
The major problem to apply Laplacian smoothing on unorganized point cloud
is the absence of the 1-ring neighbors information. Our meshless Laplacian
smoothing approach identifies the 1-ring neighbors from the unorganized point
cloud and then apply the Laplacian smoothing algorithm mentioned above.
}

In the absence of a mesh, we cannot define $N(p_i)$ to be the 1-ring neighbor
of $p_i$.  The obvious alternative is to define $N(p_i)$ to be the points in
$P''$ within some distance from $p_i$.  However, there are two problems with
this approach.  First, there can be quite a lot of points within a predefined
distance from $p_i$.  Working with all these points for every $p_i$ slows down
the computation, and according to our experience, the output does not appear to
be better.  Second, 
%even if a point $p_i$ is surrounded by other points in
the point distribution in $P''$ can be uneven and this uneven distribution may
make $p_i$ drift in the direction of higher local density.  We observe that
this may produce poorer output.

We address these two problems by building an octree for $P''$ and use the
octree to define a smaller point set $Q$ that is representative of $P''$.  The
set $Q$ has fewer points than $P$, so it improves the computational efficiency.
%Although the problem of non-uniform local density is alleviated by this fact,
We still have to define $N(q)$ for every $q \in Q$ to address the issue of
non-uniform local density.  Eventually, we forget about $P''$ and smooth $Q$
instead, i.e., we call {\sc Smooth}$(Q,\lambda,\gamma)$ repeatedly until no
point in $Q$ moves or until the number of repetitions reaches a predefined
threshold.  In the following, we describe the construction of $Q$ and $N(q)$
for every $q \in Q$.

First, construct an octree $T_{P''}$ for $P''$ as described in
Section~\ref{sec:octree}.  Second, for each leaf node $x$ of $T_{P''}$, compute
the mean point $q_x$ of the subset of $P''$ in $x$.  The set $Q$ is the
resulting collection of mean points.  By Remark~1 in Section~\ref{sec:octree},
we expect $Q$ to be a noisy locally uniform sample.

\cancel{
The smoothing can be done much more efficiently if
each leaf node contains one point only, and thus we subsample from $P''$ by
picking exactly one point from each leaf node as follows. For each leaf node
$x$ in $T_{P''}$, we compute the average point $p_x$ of $x$ by taking the
average of all sample points in $x$.
}

\begin{figure}
\centering
{\includegraphics[scale=0.475]{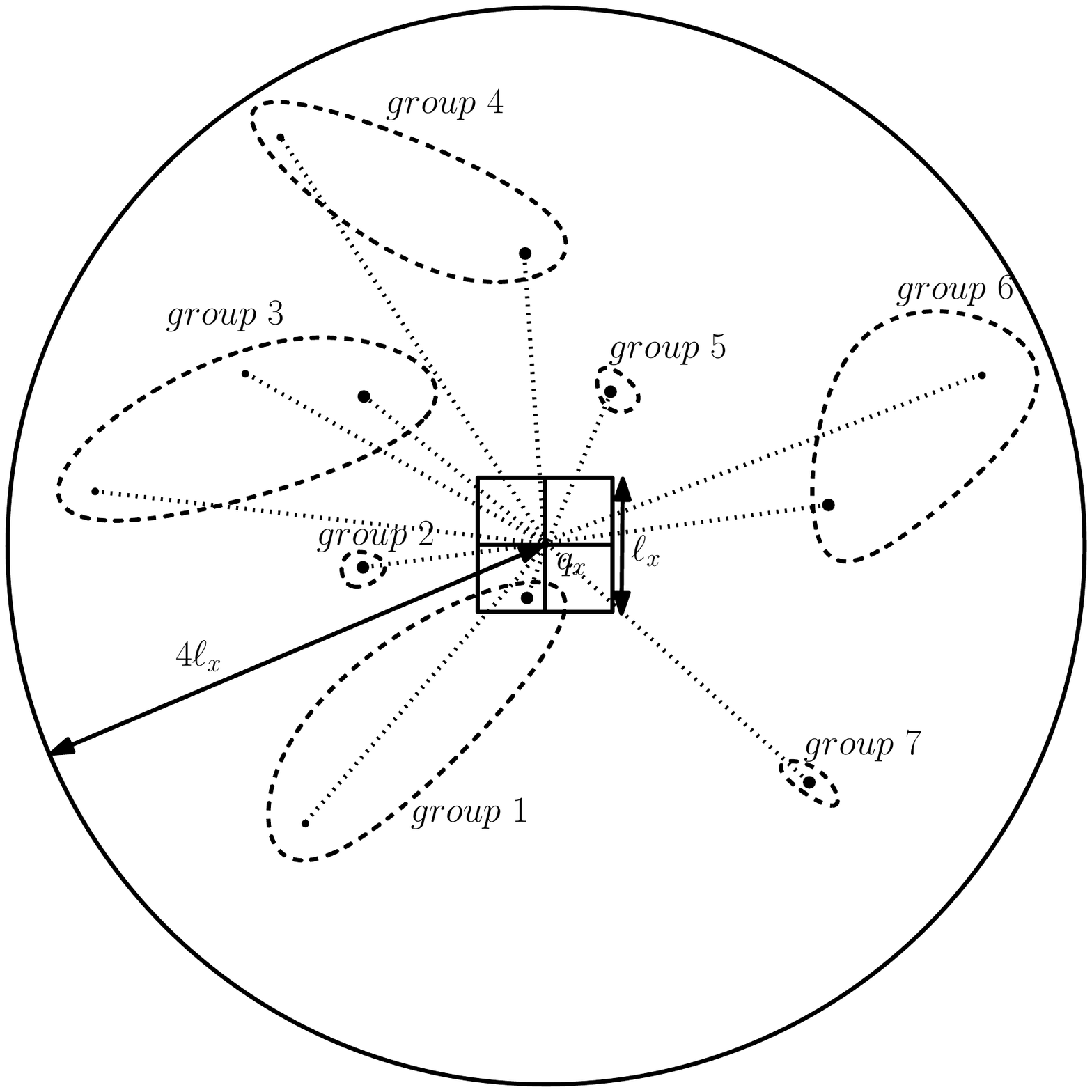}}
%\scalebox{0.8}{\includegraphics{figures/algorithm/oneRingNeighbors.pdf}}
\caption{A 2D example of the grouping process.
The boundary of the square centered at $q_x$ is divided into $8$ segments. The dotted lines 
are parts of rays from $q_x$ to points in $Q \cap B$. 
The points enclosed by the same dashed curve are in the same
group.   The thicker dots are the points in $N(q_x)$.} 
\label{fig:oneRingNeighbors}
\end{figure}

Consider a point $q_x \in Q$.  Recall that $\ell_x$ denotes the size of the
leaf node $x$ in $T_{P''}$.  Let $B$ be the ball centered at $q_x$ with radius
$4\ell_x$.  Let $C$ be the axes-aligned cube centered at $q_x$ with side length
$\ell_x$.  Partition the boundary of $C$ into 24 equal squares of side length
$\ell_x/2$.  For any two points $q_y, q_z \in Q \cap B$, we put $q_y$ and $q_z$
into the same group if the rays from $q_x$ through $q_y$ and $q_z$,
respectively, intersect the same square in the partition of the boundary of
$C$.  As a result, $Q \cap B$ are divided into at most 24 groups.  For each
group, we pick the point in the group that is closest to $q_x$, and include
this point in $N(q_x)$.   Therefore, $N(q_x)$ consists of one point from each
non-empty group.  Selecting one point per group addresses the problem of
non-uniform local density. 
\cancel{
In the computation of the 1-ring neighbors for each point $p_x$, we define two
points $p_x$ and $p_{x'}$ to be neighbors if the distance between them is at
most $4l_{x}$. Consider a cube $C$ centered at $p_x$ with side length $l_{x}$.
Divide the boundary of $C$ into $24$ squares with side length $\frac{1}{2}l_x$.
For any two neighbors $p_{x_1}$ and $p_{x_2}$ of $p_x$, let $L_1$ be the line
through $p_x$ and $p_{x_1}$, $L_2$ be the line through $p_x$ and $p_{x_2}$. Let
$s_1$ be the square on $C$'s boundary reached when we walk in the direction
from $p_x$ to $p_{x_1}$ along the line $L_1$. Let $s_2$ be the square on $C$'s
boundary reached when we walk in the direction from $p_x$ to $p_{x_2}$ along
the line $L_2$. If $s_1 = s_2$, we say $p_{x_1}$ and $p_{x_2}$ are in the same
group, otherwise they are in different groups. For each group $G$, let $p_G \in
G$ closest to $p_x$ is a 1-ring neighbor of $p_x$. By this way, the 1-ring
neighbors for each point $p_x$ would be the closest points surround $p_x$ which
approximates the 1-ring neighbors of $p_x$ in the unknown mesh for Laplacian
smoothing. 
}
Fig.~\ref{fig:oneRingNeighbors} illustrates this grouping method in 2D.
\cancel{
The square is divided into $8$ segments. The dotted lines are the lines from $p_x$
to its neighbors, the points enclosed by the same dashed line are in the same
group and the larger dots are the 1-ring neighbors of $p_x$.
}

\begin{figure}
\centering
\begin{tabular}{c}
\includegraphics[scale=0.7]{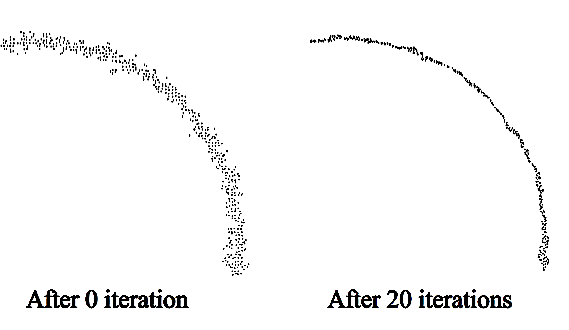} \\
\includegraphics[scale=0.7]{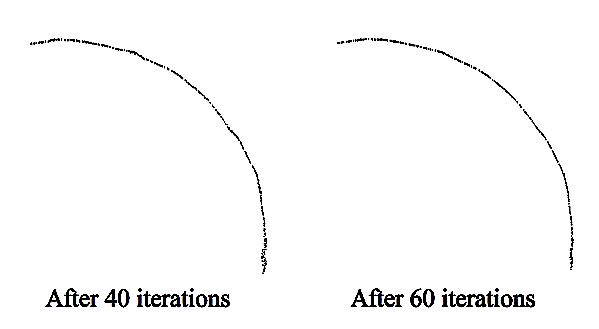} 
\end{tabular}
%{\includegraphics[scale=0.5]{figures/algorithm/meshlessLaplacianSmoothingExample.pdf}}
%\scalebox{0.8}{\includegraphics{figures/algorithm/meshlessLaplacianSmoothingExample.pdf}}
\caption{A 2D example of the meshless Laplacian smoothing effect.}
\label{fig:meshlessLaplacianSmoothingExample}
\end{figure}

Running {\sc Smooth}$(Q,\lambda,\gamma)$ repeatedly gives 
our final denoised point set.  Fig.~\ref{fig:meshlessLaplacianSmoothingExample}
shows the effect of our meshless Laplacian smoothing on the right image in
Fig.~\ref{fig:removeHighNoise}.  Only the upper right quarter of the circle is
shown.

\cancel{
After obtaining the 1-ring neighbors for each point $p_x$ in $T_{P''}$, we
apply the Laplacian smoothing algorithm introduced above to denoise the point
set. (Note that each leaf node $x$ has only one point $p_x$ now) Let $P_f$
denote the final denoised point set obtained by extracting the point $p_x$
(possibly moved during smoothing) from each leaf node $x$ in $T_{P''}$. 
Figure \ref{fig:meshlessLaplacianSmoothingExample} shows the effect of our meshless
Laplacian smoothing on the 2D circle after filtering the very noisy points and
is zoomed to the top-right quarter of the circle.
}

%\section{Surface Reconstruction}
%\label{sec:SurfaceReconstruction}
%\section{Postprocessing On The Reconstructed Surface}

\section{Postprocessing}
\label{sec:Postprocessing}

Let $P_f$ denote the final denoised point set.  
%The user can run his/her favorite surface reconstruction algorithm on $P_f$ or
%perform other quantitative analyses.  %\cite{paper:Bernardini99,
%paper:RobustCocone, paper:Kazhdan06}. 
In our experiments, we run Robust Cocone~\cite{paper:RobustCocone} on $P_f$
because no normal information is required.  

Robust Cocone fills all holes in the point cloud to produce a water-tight
surface.  So boundary holes are patched as well.  We observe that the triangles
that fill a boundary hole usually have larger circumradii because the Robust
Cocone algorithm reconstructs a surface based on Delaunay triangulation.  We
recreate the boundary holes by removing triangles according to a simple
statistical criterion.  Let $r_{\mathrm{avg}}$ be the average circumradius of
the triangles in the reconstruction.  Let $r_{\mathrm{sd}}$ be the standard
deviation of the circumradii of these triangles.  Let $\varepsilon$ be a
parameter.  (We set $\varepsilon = 10$ in our experiments.)  We remove all
triangles whose circumradii are greater than $r_{\mathrm{avg}} + \varepsilon
r_{\mathrm{sd}}$.  This step can be skipped if a water-tight surface is
desired.

The resulting triangular mesh may not be sufficiently smooth as mentioned
in~\cite{paper:RobustCocone}. Therefore, we smooth the mesh using the classic
Laplacian smoothing based on the 1-ring neighbors in the mesh.

%%%%%%%%%%%%%%%%%%%%%%%%%%%%%%%%%%%%%%%%%%%%%%%%%%%%%%%%%%%%%%%%%%%%%%%%%
%                                                                       %
% An example of a table. Note how the Table Number is generated in the  %
% list of tables.                                                       %
%                                                                       %
%%%%%%%%%%%%%%%%%%%%%%%%%%%%%%%%%%%%%%%%%%%%%%%%%%%%%%%%%%%%%%%%%%%%%%%%%

%\begin{table}[h]
%\caption{Comparitive Orders of Complexity}
%\end{table}

\section{Experiments}

We implemented our algorithm and did the experiments on a Linux platform on a
PC
%with an Intel Xeon CPU 5160 3.00GHz and 4GB RAM. 
with an AMD Opteron(tm) processor 844 (1792.513 MHz) and about 8GB free memory
available to our experiments.  We compare our code with the following 
reconstruction codes: 
%including triangulation algorithms and implicit function approaches:
\begin{itemize}

\item Method~1: Robust Cocone~\cite{paper:RobustCocone}

\item Method~2: AMLS~\cite{paper:Dey05}

%\item Method~3: PoissonPU~\cite{paper:Nagai09}

\item Method~3: Sign-the-Unsigned~\cite{paper:Mullen10}

\item Method~4: Noise-adaptive distance function~\cite{paper:Giraudot13} 

\end{itemize}
The codes for methods~1, 2, 3, and 4 were obtained from the authors.
%The code for method~3 was downloaded from the author's webpage.

Methods 1 and 2 are not catered for outliers and white noise.  Therefore, the
input to methods 1 and 2 in our experiments are obtained after removing
outliers and white noise as described in Section~\ref{sec:RemoveOutliers}.
Noisy perturbation may remain though.  Methods~1,~2, and~4 were run on the same
machine as our code.  
%Methods~3 and~4 were run on the Window 7 platoform on a PC with an Intel Core2
%duo CPU E8400 3.00GHz and 4GB RAM because the executables of PoissonPU and
%Sign-the-Unsigned run on Window platform.  
Method~3 was run on the Window 7 platform on a PC with an Intel Core2 duo CPU
E8400 3.00GHz and 4GB RAM because the executable of Sign-the-Unsigned runs on
Window platform.  We are mostly concerned with comparing the noise handling
capability of our code with the other codes. 
%The running times of our code is very small.  However, 
Although the execution times of our code are almost always the smallest in the
experiments, it is not meaningful to compare execution times because the
platforms may be different, the codes may not be optimized fully, and the codes
may not be compiled with the best possible options for the respective computing
environments.  Thus, we only show our execution times but not the others.

Due to storage constraint, all images in Figures~7--16 are not included
in this document.  They can be found at 
https://www.cse.ust.hk/faculty/scheng/self-paper-short-images.pdf.

%
%As a result, it is not too meaningful
%to compare the running times of our code with Methods~3 and~4. 

%
%the running times of our
%code and PoissonPU reported later cannot be compared.

\subsection{Data Sets}
\label{sec:DataSets}

%Armad, Chair, Dragon, Buddha, Rame: 0, 0.5, 1, 2

We used $7$ data sets: Armadillo, Bunny, Chair, Dragon, 
%Frog, 
Happy Buddha, 
%Isidore Horse (I.H.), 
Ramesses, and wFish.  Bunny, Chair, Dragon, 
%Isidore Horse, 
Ramesses, and wFish
are raw point sets.  In particular, the Dragon data set contains sample points
from the background.  Armadillo
%, Frog, 
and Happy Buddha consist of vertices of reconstructed meshes. 

To each point in each data set, we add a Gaussian noise with mean zero and
standard deviation $x$ in a direction chosen uniformly at random.  There are
four different settings of $x$: 0\%, 0.5\%, 1\% and 2\% of the diagonal length
$D$ of the minimum axes-aligned bounding box of the data set.  
%That is, each
%data set is experimented with $4$ different noise levels. 

To each data set, 5000 white noise points are picked from a uniform
distribution in the minimum axes-aligned bounding box of the data set. Each white
noise point $p$ has a probability 0.05 to become a cluster of outliers if there
is no sample point inside the ball centered at $p$ with radius $0.05D$.  Let
$r_1$ be an integer chosen uniformly at random from $[1,R]$, where $R$ is some
integer fixed \emph{a priori}.  Let $r_2$ be a real number chosen uniformly at
random from $[0,1]$. If $p$ is chosen to be a cluster of outliers, then we pick
$r_1$ outlier points uniformly at random inside the ball centered at $p$ with
radius $0.001 r_2 D$.

The first column in Fig.~7 shows two noisy input data sets
with 5000 white noise points, outliers generated as described above with $R =
400$, and 1\% Gaussian noise.
%These point sets are the input for our code and method~5.  
The second column shows the point sets
after removing the white noise and outliers. 
%and these point sets are input for the methods 1--4.
The third column shows the final denoised
point sets produced by our denoising code.  
%Dragon is a raw dataset, so some data points are sampled from the flat
%background from different views.

To push the codes to the extreme, we also experimented with data sets
contaminated with 60\%, 80\% and 100\% white noise points followed by random
outlier cluster generation as discussed above (no Gaussian noise is added).   
Fig.~8 shows two examples.

%PoissonPU requires the input points to be associated with the surface normals
%which do not come with the datasets that we experimented with.  We use the
%normal estimation routine in the AMLS algorithm to estimate the surface normals
%at the input points, and then we feed these normals to PoissonPU. 

\subsection{Parameters}
\label{sec:ParametersSetting}

As discussed in Section~\ref{sec:denoise}, our algorithm requires several
parameters $\alpha,\beta,\lambda,\gamma$, $\epsilon$, and the threshold on the
number of calls to {\sc Smooth}$(Q,\lambda,\gamma)$. 
In our experiments, we set $\alpha = 2, \beta = 2, \lambda = 0.25, \gamma = 40$
and $\epsilon = 10$, but we change $\alpha$ and $\lambda$ for Happy Buddha, the
Tortile column, and Pierre's clenched fist.  

The larger $\alpha$ is, the larger the octree leaf nodes
(Section~\ref{sec:octree}).  Larger leaf nodes are needed when the sampling is
significantly non-uniform at some places.  
%The setting of $\alpha = 2$ works for almost all of the data sets that we
%tried.  
We increase $\alpha$ to 4 for Pierre's clenched fist due to the significantly
non-uniform sampling (Fig.~9(b)).   When we work with data sets
with 60\%, 80\% and 100\% white noise points, the setting of $\alpha = 2$ still
works for almost all data sets except that we need to set $\alpha = 2.5$ for
Happy Buddha.

The larger $\lambda$ is, the greater the step size in Laplacian smoothing,
which gives a greater smoothing effect.  The setting of $\lambda = 0.25$ works
for the Gaussian noise levels of 0\%, 0.5\%, 1\%, and 2\% that we tried.  If it
is known that the noise level is small, say less than 0.25\%, then $\lambda$
can be reduced and the effect is that more features can be preserved.  For the
Tortile column, the setting of $\lambda = 0.1$ preserves more features than the
setting of $\lambda = 0.25$ (Fig.~9(a)).

\cancel{
 for all the models except the Happy Buddha. We change $\alpha$
from $2$ to $4$ for the Happy Buddha model because it contains some significant
non-uniform regions. The effect is that the outlier removal step removes more
points near the real surface. This shows that, by changing the parameters, our
algorithm can also handle non-uniform point sets.
}

The threshold on how many times {\sc Smooth}$(Q,\lambda,\gamma)$ is called is
determined before smoothing.  Translate and scale $Q$ so that a minimum
axes-aligned bounding cube of $Q$ centered at the origin has side length $2$.
For each point $q \in Q$, let $d_q$ denotes the average distance between $q$
and $N(q)$ (Section~\ref{sec:MeshlessLaplacianSmoothing}).  Let
$d_{\mathrm{avg}}$ be the average of $d_q$ over all $q \in Q$.  Define $L =
\lfloor (d_{\mathrm{avg}}^2\cdot |Q|)/2 \rfloor$ to be the threshold desired,
which works well in our experiments.
%From our experiment, $L$ gives a good limitation on the number of iterations.
%Table~\ref{table:maximumNumberOfIteration} shows the number of iterations for
%our datasets.  each models with the $4$ different noise levels that we have
%experimented.

%PoissonPU requires several parameters. One of them is the approximation error
%tolerance. As mentioned in~\cite{paper:Nagai09}, this parameter should be small
%when the noise level is low (to preserve more features) and it should be large
%when the noise level is high (to produce a smooth surface).  The default value
%is $0.01$.  After some tuning, it seems best to set this parameter to 0.001,
%0.025, 0.035 and 0.07 for the noise levels of 0\%, 0.5\%, 1\%, and 2\%
%respectively.
%% for 
%%$0\%$ sd Gaussian noise, $0.025$ for $0.5\%$ sd Gaussian noise, $0.035$ for
%%$1\%$ sd Gaussian noise and $0.07$ for $2\%$ sd Gaussian noise in our
%%experiments. 
%Another parameter controls the number of points in the resulting mesh, and the
%default value is $200$.  We set this parameter to $400$ in order to show more
%details.

Sign-the-Unsigned has a maximum depth level parameter.  A higher depth level
means finer level of details in the output, possibly at the expense of
a higher running time.  We set the maximum depth level to be 10.
The depth levels used in~\cite{paper:Mullen10} are between 8 and
12.

\subsection{Experimental Results and Comparison}

Consider the experiments with 5000 white noise points, randomly generated
clusters of outliers, and $x$\% Gaussian noise, where $x \in \{0,0.5,1,2\}$.
Fig.~10--14 show some
reconstructions by our code and methods 1--4.  Tables~\ref{table:times} gives
the statistics of these experiments.  Recall that the input to Robust Cocone
and AMLS has been subject to white noise and outlier removal using our code
(noisy perturbation may remain).   Noise-Adaptive seems to have difficulty in
handling the white noise and clustered outliers in our experiments, and it
produced corrupted output.

\begin{table*}
\caption{
%The input size includes the
%number of points in the original dataset, the white noise
%points, and the outliers.  
Each data set is contaminated with 5000 white noise points, randomly
generated outlier clusters, and Gaussian noise.  The input size includes
the perturbed data points, white noise, and outliers.  The column
``Denoising Time'' gives the execution time of our denoising step.  The column
``Total Time'' is the total running time including the denoising step and
the surface reconstruction step.  
Our code, Sign-the-Unsigned, and Noise-Adaptive were run on the same contaminated 
data sets, but Robust Cocone and AMLS were run on point sets that have 
been subject to white noise and outlier removal by our code.  A cell is marked $\times$
if the corresponding reconstruction is corrupted or unsuccessful.  Otherwise,
the cell is marked $\surd$.} 
\label{table:times}
\centerline{
\begin{tabular}{|l|r|r|r|r||r|r|r|r|r|}
\hline
& Gaussian & Input & Denoising & Total & Our & Robust & AMLS & Sign-the- & Noise- \\
& Noise Level & size  & Time   & Time  & Code & Cocone &      & Unsigned  & Adaptive \\
\hline
\multirow{4}{*}{Armadillo}
& 0\% & 197640 & 14.36s & 132.51s & $\surd$ & $\surd$ & $\surd$ & $\surd$ & $\times$ \\
\cline{2-10}
& 0.5\% & 199518 & 29.38s & 113.70s & $\surd$ & $\times$ & $\times$ & $\surd$ & $\times$ \\
\cline{2-10}
& 1\% & 199117 &  44.48s & 112.99s & $\surd$ & $\times$ & $\times$ & $\surd$ & $\times$ \\
\cline{2-10}
& 2\% & 199002 & 45.71s & 84.33s & $\surd$ & $\times$ & $\times$ & $\times$ & $\times$ \\
\cline{1-10}
\multirow{4}{*}{Bunny}
& 0\% & 385113 & 38.62s & 346.24& $\surd$ & $\surd$ & $\surd$ & $\surd$ & $\surd$ \\
\cline{2-10}
& 0.5\% & 392522 & 91.51s & 240.89s & $\surd$ & $\surd$ & $\times$ & $\surd$ & $\times$  \\
\cline{2-10}
& 1\%   & 380930 & 125.29s & 270.58s & $\surd$ & $\times$ & $\times$ & $\surd$ & $\times$  \\
\cline{2-10}
& 2\%   & 382294 & 145.87s & 290.91s & $\surd$ & $\times$ & $\times$ & $\surd$ & $\times$ \\
\cline{1-10}
\multirow{4}{*}{Ramesses}
& 0\% & 824488 & 54.87s & 1055.11s & $\surd$ & $\surd$ & $\surd$ & $\surd$ & $\times$ \\ 
\cline{2-10}
& 0.5\% & 821297 & 143.45s & 591.97s & $\surd$ & $\surd$ & $\surd$ & $\surd$ & $\times$ \\  
\cline{2-10}
& 1\% & 824296 & 161.52s & 597.21s & $\surd$ & $\surd$ & $\times$ & $\surd$ & $\times$ \\ 
\cline{2-10}
& 2\% & 825584 & 176.78s & 389.87s & $\surd$ & $\surd$ & $\times$ & $\surd$ & $\times$ \\ 
\cline{1-10}
\multirow{4}{*}{Chair}
& 0\% & 246042 & 16.48s & 242.10s & $\surd$ & $\surd$ & $\surd$ & $\times$ & $\times$ \\ 
\cline{2-10}
& 0.5\% & 242940 & 28.74s & 91.03s & $\surd$ & $\times$ & $\times$ & $\times$ & $\times$ \\
\cline{2-10}
& 1\% & 252343 & 38.64s & 105.36s & $\surd$ & $\times$ & $\times$ & $\times$ & $\times$ \\ 
\cline{2-10}
& 2\% & 242693 & 45.00s & 113.00s & $\surd$ & $\times$ & $\times$ & $\times$ & $\times$ \\ 
\cline{1-10}
\multirow{4}{*}{Dragon}
& 0\% & 2767387 & 351.39s & 3512.06s & $\surd$ & $\surd$ & $\surd$ & $\times$ & $\times$ \\ 
\cline{2-10}
& 0.5\% & 2767110 & 1018.29s & 3537.03s & $\times$ & $\times$ & $\times$ & $\times$ & $\times$ \\  
\cline{2-10}
& 1\% & 2767461 & 1030.59s & 3690.72s & $\times$ & $\times$ & $\times$ & $\times$ & $\times$ \\ 
\cline{2-10}
& 2\% & 2768586 & 1401.62s & 3026.28s & $\times$ & $\times$ & $\times$ & $\times$ & $\times$ \\ 
\cline{1-10}
%\multirow{4}{*}{Frog} 
%& 0\% & 225K & 197K & 129K & 10s & 123s & 279s & 1012s & 284s+28s	  \\
%\cline{2-10}
%& 0.5\% & 218K & 191K & 160K & 21s & 93s & 33s & 791s & 41s+12s	  \\
%\cline{2-10}
%& 1\% & 219K & 197K & 102K & 26s & 53s & 35s & 958s & 44s+14s	  \\
%\cline{2-10}
%& 2\% & 218K & 195K & 124K & 39s & 71s & 42s & \cellcolor[gray]{0.8} 1537s &	45s+13s  \\
%\cline{1-10}
%\multirow{2}{*}{Happy} 
& 0\% & 590779 & 36.72s & 605.94s & $\surd$ & $\surd$ & $\times$ & $\times$ & $\times$ \\ 
\cline{2-10}
Happy & 0.5\% & 583277 & 89.37s & 412.26s & $\surd$ & $\times$ & $\times$ & $\times$ & $\times$ \\
\cline{2-10}
%\multirow{2}{*}{Buddha} 
Buddha
& 1\% & 588542 & 111.55s & 247.29s & $\times$ & $\times$ & $\times$ & $\times$ & $\times$ \\ 
\cline{2-10}
& 2\% & 587481 & 137.32s & 298.02s & $\times$ & $\times$ & $\times$ & $\times$ & $\times$ \\ 
\cline{1-10}
%\multirow{4}{*}{I.H.} 
%& 0\% & 1.10M & 1.07M & 502K & 35s & 382s & 1003s & \cellcolor[gray]{0.8} 8256s & \cellcolor[gray]{0.8} 447s+86s	  \\
%\cline{2-10}
%& 0.5\% & 1.10M & 1.00M & 818K & 132s & 532s & 604s & \cellcolor[gray]{0.8} 18173s &	\cellcolor[gray]{0.8} 280s+31s  \\
%\cline{2-10}
%& 1\% & 1.10M & 1.06M & 555K & 157s & 321s & 770s & \cellcolor[gray]{0.8} 38140s & \cellcolor[gray]{0.8} 302s+40s	  \\
%\cline{2-10}
%& 2\% & 1.09M & 1.05M & 673K & 238s & 476s & 852s & \cellcolor[gray]{0.8} 84915s &	\cellcolor[gray]{0.8} 302s+59s  \\
%\cline{1-10}
\multirow{4}{*}{wFish} 
& 0\% & 303478 & 14.87s & 122.41s & $\surd$ & $\surd$ & $\surd$ & $\surd$ & $\times$ \\ 
\cline{2-10}
& 0.5\% & 306185 & 43.28s & 144.17s & $\surd$ & $\times$ & $\surd$ & $\times$ & $\times$ \\ 
\cline{2-10}
& 1\% & 304348 & 62.88s & 157.26s & $\surd$ & $\times$ & $\times$ & $\surd$ & $\times$ \\ 
\cline{2-10}
& 2\% & 303475 & 75.95s & 170.97s & $\surd$ & $\surd$ & $\times$ & $\surd$ & $\times$ \\ 
\cline{1-10}
\end{tabular}
}
\end{table*}

In the experiments with 0\% Gaussian noise, Robust Cocone may preserve more
detailed features as seen in the Armadillo and Ramesses data sets
(Figs.~10 and~11), but it may
give corrupted output as seen in the Dragon data set.  Our code, AMLS, and
Sign-the-Unsigned also preserve a lot of detailed features.  But AMLS and
Sign-the-Unsigned may lose some detailed features as seen in the Ramesses data
set.  Sign-the-Unsigned had difficulty with the structured and clustered
outliers in the raw Dragon data set (Fig.~13).

The Chair and Happy Buddha data sets are challenging.  Refer to
Fig.~12.  In the experiments with 0\% Gaussian noise in the
Chair data set, AMLS produced a rough surface, and Sign-the-Unsigned
produced wrong topology.  In the experiments with 0.5\%, 1\% and 2\% Gaussian
noise in the Chair data set, Robust Cocone, AMLS, and Sign-the-Unsigned
produced corrupted output.  Our code still recovered the topology correctly and
the surface quality degrades gradually as the Gaussian noise increases.
Similar trends are observed in the experiments with the Happy Buddha data set
(Fig.~14), except that our code produced wrong
topology at 1\% and 2\% Gaussian noise.  The Dragon data set is also difficult
for our code (Fig.~13).  The output is already corrupted
at 0.5\% Gaussian noise (in the sense that the neck is merged with the back).

The output of our code, Robust Cocone and AMLS show that it is insufficient to
remove white noise and outliers alone.  (The input to Robust Cocone and AMLS
has been subject to removal of white noise and outliers by our code.) The
remaining noisy perturbation is still so large that Robust Cocone and AMLS
produced corrupted output. 

Our code shows a better noise handling ability in the above experiments.  Its
execution times are reasonably small as seen in Table~\ref{table:times}.
Sign-the-Unsigned is also quite robust in the above experiments.  Therefore, we
push our code and Sign-the-Unsigned to the extreme by introducing a lot more
white noise in the another set of experiments.  To each of the data sets
Armadillo, Bunny, Chair, Dragon, Happy Buddha, Ramesses, and wFish, we
introduce 60\%, 80\%, and 100\% uniformly distributed white noise points.
Then, we probabilistically turn white noise points into clusters of outliers as
described in Section~\ref{sec:DataSets}.  No Gaussian noise is added.
Sign-the-Unsigned could only produce useful output for Armadillo and Bunny in
the experiments with 60\% white noise points.  Fig.~15
shows these two output surfaces.  Our code handled these contaminated data sets
quite well.  Fig.~16 shows some of our reconstructions in
comparison with those obtained in the first set of experiments when 5000 white
noise points were introduced.  The white noise and outliers may cause too many
point deletions so that some areas become under-sampled.  This will lead to a
rough surface and/or holes.  For example, there are dents in the base of the
Happy Buddha, and there are some holes in Ramesses and wFish.

\section{Conclusion}

We propose a fast and simple algorithm to denoise an unorganized point cloud
for surface reconstruction.  It can efficiently handle a large amount of white
noise, outliers, and fairly large noisy perturbation.  In particular, we can
also filter outliers that are clustered together, which may happen to be sample
points from the background during the scanning of an object in the foreground.

There are two open research problems. One is to determine the number of
surfaces automatically. Another problem is handle sharp features and
non-manifold features.

%In the future, we can develop an approach to determine the number of surfaces which is represented by the given point set rather than specified by the users. Our algorithm can also be improved by filtering the unnecessary 1-ring neighbors in the Meshless Laplacian Smoothing step rather than use all the 1-ring neighbors, and can be improved by further modification of the Laplacian Smoothing algorithm for our denoising purpose. We may also develop a triangulation algorithm specifically for our denoised point set rather than use Robust Cocone algorithm. 

\cancel{
\begin{figure}
\centering
\begin{tabular}{c}
\scalebox{0.55}{\includegraphics{figures/experimentalResults/hole-bunny.png}} \\ 
%\scalebox{0.55}{\includegraphics{figures/experimentalResults/hole-frog.png}} 
%\scalebox{0.7}{\includegraphics{figures/experimentalResults/hole-dragon.png}} 
\end{tabular}
\caption{Retrieved boundaries.}
\label{fig:hole}
\end{figure}
}

\section*{Acknowledgment}

The authors would like to thank Jiongxin Jin for helpful discussion and the
authors
of~\cite{paper:RobustCocone,paper:Dey05,paper:Mullen10,paper:Giraudot13} for
providing their codes.

% Can use something like this to put references on a page
% by themselves when using endfloat and the captionsoff option.

\bibliographystyle{plain}

\bibliography{myThesis}

\appendix

\section*{Appendix: Images}

Due to storage constraint, the images are not included in this document.  Please
find them at \\
https://www.cse.ust.hk/faculty/scheng/self-paper-short-images.pdf.

\end{document}